\documentclass{jfm}
\usepackage{graphicx}
\usepackage{epstopdf,epsfig}
\usepackage{amsmath}
\usepackage{url}
\usepackage{graphicx}
\usepackage{xcolor}

\shorttitle{Quasigeostrophy against the wall}
\shortauthor{A. Venaille}

\title{Quasigeostrophy against the wall}

\author{A. Venaille}

\affiliation{Univ Lyon, Ens de Lyon, Univ Claude Bernard, CNRS, Laboratoire de Physique, F-69342 Lyon, France}

\date{\today} 

\begin{document}

\maketitle

\begin{abstract}
Coasts are obstructions to the classical derivation of continuously stratified quasigeostrophic equations, due to  possible resonances  between slow internal coastally trapped Kelvin waves and anticyclones.  [Deremble \textit{et al} Ocean Modelling 2017] proposed a coupled model between a quasigeostrophic interior and boundary layer Kelvin wave dynamics. We revisit the derivation of this model, paying particular attention to conservation laws. We find that quasigeostrophic energy is conserved  despite the existence of Kelvin wave shocks in the boundary layer. The effect of those  shocks is  to change the global distribution of potential vorticity, and, consequently the interior flow structure. In that respect, {\color{black} we show that} there is an active control of the boundary region on the interior flow.
\end{abstract}


\section{Introduction}

 Quasigeostrophic models  play a prominent role in our understanding of midlatitude atmospheric and  oceanic dynamics \citep{vallis2017atmospheric}. They describe the slow evolution of geostrophically balanced motion, filtering out the fast dynamics of inertia-gravity waves. Yet, in a series of recent papers, Dewar and collaborators showed that  geostrophically balanced motion in continuously stratified fluid may interact with slow internal Kelvin waves trapped  along a lateral wall, {\color{black} and this is} an obstruction to the classical derivation of quasigeostrophic equations: \cite{dewar2010topographic} identified a mechanism of potential vorticity  injection in interior flows through the formation Kelvin wave shocks; \cite{dewar2011submesoscale} addressed the relevance of this process within the oceanic energy cycle; \cite{hogg2011kelvin} deciphered how and when Kelvin wave shocks are generated by an initially geostrophic flow, following previous work on rotating hydraulics \citep{pratt2007rotating}. Building upon these results, \cite{deremble2017coupled} proposed a coupled model between interior quasigeostrophic dynamics and a boundary layer equation describing nonlinear Kelvin wave dynamics.  {\color{black} The so-called Deremble-Johnson-Dewar} model captured the generation of cyclones by shocks following the impact of an anticyclone on a coast. This mechanism of potential vorticity generation by shocks bears similarities with rip-current formation in the surf zone \citep{peregrine1998surf,buhler2000vorticity}, albeit at a different scale. The main difference here is that Kelvin wave shocks only produce cyclones. \cite{deremble2017coupled} emphasised the key role of this boundary layer dynamics in shaping {\color{black} the} interior flow properties close to the wall.  They also 
 {\color{black} found via numerical simulations} that this process acts as a significant sink of energy, but without providing scaling with respect to the Rossby number, the small parameter of the asymptotic model.
 
The aim of this paper is to clarify how global conservation laws of standard,  unbounded quasigeostrophic models are affected by the presence of a coast, by revisiting the derivation of \cite{deremble2017coupled}. The paper is organised as follows. We introduce in a second section the hydrostatic, rotating Boussinesq equations, and we explain why the presence of a wall makes the derivation of quasigeostrophic equations difficult. Starting from the multiple layer shallow water model with sufficiently thin layer thickness, a new derivation of Deremble-Johnson-Dewar model is proposed in a third section, paying particular attention to mass conservation, energy conservation, and a local model for potential vorticity injection by shocks. We end in a fourth section with a discussion on symmetries and on possible geophysical applications.

\section{Boussinesq syllabus}
\subsection{Hydrostatic Boussinesq dynamics on the f-plane}

Our starting point is the 3D Boussinesq, hydrostatic equations with traditional approximation for the Coriolis force \citep{vallis2017atmospheric}. This is a standard model for geophysical flows, including the effect of rotation and stratification through the Coriolis parameter $f$ (twice the projection of the planet rotation rate on the local vertical axis) and the buoyancy frequency $N$. 
Calling $L$ and $H$ the typical horizontal and vertical length scales of the flow with typical velocity $U$, the Boussinesq dynamics admits three non-dimensional parameters: the aspect ratio, the Rossby number and the Burger number, defined as
\begin{equation}
 \alpha\equiv  \frac{H}{L},
 \quad \mathrm{Ro} \equiv\frac{U}{Lf},\quad \mathrm{Bu}\equiv  \left(\frac{NH}{fL}\right)^2 .
\end{equation}
The hydrostatic limit corresponds to $\alpha\ll 1$. The hydrostatic Boussinesq system is 
\begin{eqnarray}
\partial_x u  +\partial_y v+\partial_z w &=& 0\\
0 &=&-\partial_z p +b\\
 \mathrm{Ro} \left(\partial_t  +u\partial_x  +v\partial_y+w\partial_z  \right)u&=&-\partial_x p +v\\
\mathrm{Ro} \left(\partial_t  +u\partial_x  +v\partial_y+w\partial_z  \right)v &=&-\partial_y p -u\\
\mathrm{Ro}\left(\partial_t  +u\partial_x  +v\partial_y+w\partial_z  \right)b  &=&-\mathrm{Bu} w  
\end{eqnarray}
The field $b$ is the perturbation buoyancy corresponding to rescaled density anomalies around the stable stratification. To simplify the discussion, we consider the case where $f$ and $N$ are constant.

\subsection{Plane waves} 

\begin{figure}
    \centering
    \includegraphics[width=\columnwidth]{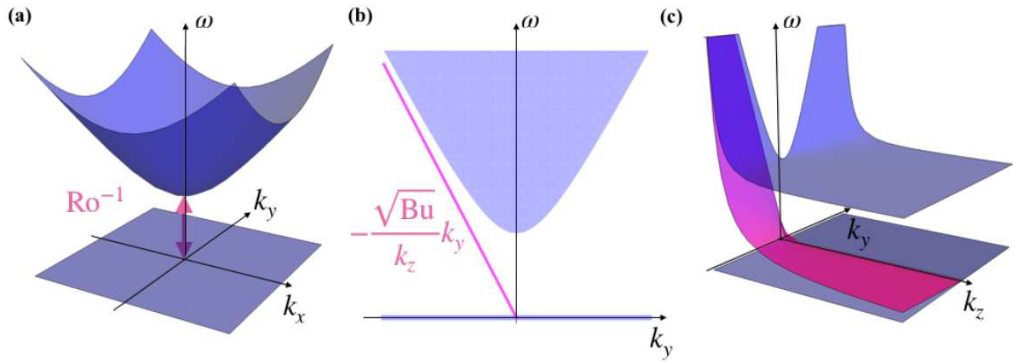}
    \caption{Dispersion relation of hydrostatic Boussinesq model linearized around a state of rest{, \color{black} adapted from \citep{zeitlin2018geophysical}}. (a) Unbounded case, $k_z$ fixed. There is a frequency gap $\Delta \omega=f$ between the inertia-gravity wave band and the (flat) geostrophic wave band. (b) Coastal case (wall at $x=0$), $k_z$ fixed. The dispersion relation for varying values of $k_x$ is projected in $k_y,\omega$ plane. The magenta line corresponds to coastally trapped internal Kelvin waves. (c) Coastal case, $k_z>0$. The blue surface at the top represents the bottom boundary of the inertia-gravity wave band. The magenta surface corresponds to Kelvin-wave dispersion relation. When $k_z$ tends to $\infty$ for a given value of $k_y$, the Kelvin wave dispersion relation tends to the flat geostrophic band. This is an obstruction to classical derivation of quasigeostrophic equations.}
    \label{fig:linear_disp}
\end{figure}

We first consider a case without boundary, and look for solutions of the hydrostatic Boussinesq equations linearized around a state of rest. Eigenmodes are on the form $ e^{i\omega t -ik_x x-ik_y y-ik_z z}$, and the problem admits three wave bands with dispersion relations
\begin{equation}
    \omega=\pm \frac{1}{\mathrm{Ro}} \sqrt{ 1+\frac{\mathrm{Bu}}{k_z^2} \left(k_x^2+k_y^2\right)},\quad \omega =0 \ .\label{eq:disp_bulk}
\end{equation}
For a given $k_z$, we recover the dispersion relation of shallow water waves with celerity $c=N/|k_z|$, see figure \ref{fig:linear_disp}. The zero frequency wave band corresponds to geostrophic modes, for which the pressure force balance the Coriolis force. The non-zero frequency bands corresponds to hydrostatic, internal inertia-gravity waves. Geostrophic modes and inertia-gravity wave modes are separated by a frequency gap of width $\mathrm{Ro}^{-1}$. The existence of this gap is central to the classical derivation of the quasigeostrophic model.

\subsection{Unbounded quasigeostrophic dynamics}

The  dynamics of geostrophic modes can be decoupled from the  dynamics of internal gravity waves modes in the small Rossby number limit $\mathrm{Ro} \ll 1$. This amounts to consider a wide frequency gap limit between (slow) geostrophic and (fast) internal inertia-gravity wave modes. The quasigeostrophic model describes the slow dynamics of geostrophic mode. It is derived  through an asymptotic expansion, with a small parameter given by the Rossby number $Ro\ll 1$, for a fixed Burger number $Bu \sim 1$ \citep{vallis2017atmospheric}. This last condition means that typical horizontal flow structures $L$ are of the order of an intrinsic length scale named, the Rossby radius of deformation $NH/f$.

\subsection{Internal coastal Kelvin waves}

The presence of a lateral wall allows for the along-wall propagation of a new class of  waves trapped in the across-wall direction, with frequencies filling the frequency gap between inertia-gravity waves and geostrophic modes. Those are the celebrated internal coastal Kelvin waves. Their salient features are derived from the hydrostatic Boussinesq model linearized around a state of rest, in the presence of a lateral (vertical) wall along the $y$-direction{\color{black}. We take the wall at $x=0$ and consider the}  flow taking place in the region $x>0$, with impermeability boundary condition at the wall: $u(0,y,z,t)=0$. Eigenmodes are of the form $g(x) e^{i \omega t - i k_y y - i k_z z} $ with $g(x)$ to be determined. There is two classes of eigenmodes. First, the bulk modes, with $g(x)=\sin(k_x x)$ and with the same dispersion relation as in the unbounded case (\ref{eq:disp_bulk}). Second, an additional branch of boundary modes that correspond to internal coastal Kelvin waves, satisfying geostrophic balance in the along-wall direction with vanishing velocity in the across-wall direction:
\begin{equation}
    v=\partial_x p,\quad \text{with } \partial_{xx}p=-\mathrm{Bu}^{-1}\partial_{zz} p,\quad u=0 .\label{eq:kelvin_lin}
\end{equation} Those modes have several features that will play an central role in the derivation of a model coupling interior and boundary dynamics: they are trapped along the wall, unidirectional, and propagate as non-rotating  hydrostatic internal gravity waves:
   \begin{equation}
   g(x)=e^{-x/l}, \quad l=\frac{\mathrm{Bu}^{1/2}}{|k_z|},\quad \omega=-\frac{\mathrm{Bu}^{1/2}}{\mathrm{Ro}} \frac{k_y}{|k_z|} \ .\label{eq:l_omega}
   \end{equation}
   Both the trapping length scale and the  phase speed vanish for large vertical wavenumbers.

We readily see on the dispersion relation plotted  figure \ref{fig:linear_disp} that the presence of a new branch of Kelvin wave modes filling the frequency gap is an obstruction to the classical derivation of quasigeostrophic dynamics: whatever the value of the horizontal wave number $k_x$ and the value of the frequency $\omega$, there is a value of vertical wavenumber $k_z$ such that a coastal wave exists.  
This means that one can not dismiss the presence of coastal waves when performing the standard multiple scale expansions leading to quasigeostrophic dynamics. 

{\color{black} Let us consider a Kelvin waves with wavenumber $k_y\sim 1$ and frequency $\omega$. Interactions between this wave and geostrophic modes having a typical eddy turnovertime $L/U\sim 1$ occur when $\omega \sim 1$. Injecting this scaling in Eq. (\ref{eq:l_omega}) and assuming $\mathrm{Bu}\sim 1$ leads to $1/k_z \sim \mathrm{Ro}$ and $l\sim \mathrm{Ro}$ (with dimensions, this gives  $1/k_z^*\sim   \mathrm{Ro} H$ and $l^*\sim \mathrm{Ro} L$).}  To conclude, the linear analysis offers important physical insights on possible coupling between {\color{black} the} bulk (interior) geostrophic modes and {\color{black} the} boundary (coastally trapped) Kelvin waves in the limit of vanishing Rossby numbers, with three properties that will be essential features of Deremble-Johnson-Dewar model: 
\begin{itemize}
    \item The interactions involves  internal Kelvin waves having a vertical wavelength that scales linearly with the Rossby number, and being confined in a boundary layer with a thickness that also scales linearly with the Rossby numbers.
    \item For all the coastal Kelvin waves, the across wall velocity $u$ is identically zero. This property will hold {\color{black} at lowest order in the amplitude of the wave} for a superposition of coastally trapped modes interacting  nonlinearly{\color{black}, since the nonlinear term in the evolution of $u$ is proportional to $u$}.
    \item Since there is only one coastally trapped mode for a given value of $(k_y,k_z)$, the boundary layer dynamics that  describes the nonlinear evolution of these coastally trapped waves will be governed by a 2D equations in the $(y,z)$ plane.
\end{itemize}
\section{Coupling a quasigeostrophic interior  to Kelvin wave dynamics}

 We now revisit the derivation of {\color{black} the Deremble-Johnson-Dewar model that couples} an interior quasigeostrophic flow to  boundary layer Kelvin wave dynamics in the limit $\mathrm{Ro}\rightarrow0$ with $\mathrm{Bu}\sim 1$. While their derivation {\color{black} was} performed in the continuously stratified case with isopycnal coordinates, our starting point is the multiple layer shallow water. This is the natural discretisation of isopycnal hydrostatic Boussinesq equations, keeping track of the layerwise potential vorticity conservation. From a practical point of view, this makes direct connections with numerical simulations that deal with discretised models.  From a fundamental or pedagogical perspective, this makes possible a direct application of previous results on rotating shallow water hydraulics \citep{pratt2007rotating,zeitlin2018geophysical}. The continuous case {\color{black} in density coordinate} is recovered in the limit of vanishing layer thickness.  

The multiple-layer shallow water model is written as a triplet of dynamical equations for each layer $i$ (with $i$ increasing upward) with depth independent horizontal  velocity $\mathbf{u}_i=(u_i,v_i)$ and thickness $h(1+\delta \eta_i)$:
\begin{eqnarray}
 \mathrm{Ro}\left(\partial_t +\mathbf{u}_i\cdot \nabla  \right)u_i   &=&-\partial_x p_i +v_i \label{eq:ui_start}\\
  \mathrm{Ro}\left( \partial_t +\mathbf{u}_i \cdot \nabla \right)v_i    &=&-\partial_y p_i - u_i \label{eq:vi_start}\\
\left(\partial_t +\mathbf{u}_i\cdot\nabla\right) \delta \eta_i &=&-\left(1+\delta\eta_i\right)\nabla\cdot \mathbf{u}_i \label{eq:hi_start}
\end{eqnarray}
which express momentum conservation and mass conservation, respectively.  
Interface thickness variations and pressure fields are related  through hydrostatic balance {\color{black} (see Appendix A)}:
\begin{eqnarray}
\delta \eta_i&=&-\frac{\mathrm{Ro}}{\mathrm{Bu}}  \frac{p_{i+1}-2p_i+p_{i-1}}{\left(\delta z\right)^2},\quad \delta z =\frac{h}{H} , \label{eq:hi_pi}
\end{eqnarray}
with a constant density jump  $\Delta \rho/\rho_0$ between adjacent layers, such that $g\Delta \rho/(\rho_0 h)=N^2$. Realistic configurations would also require specific equations for the upper and lower layers (interpreted as upper and lower boundary conditions in the continuous limit); we focus here on internal layers to simplify the discussion, assuming that the domain is unbounded in the vertical (the layers are then indexed by $i\in\mathbb{Z}$). 
The mean interface thickness must be chosen sufficiently thin  to allow for possible resonances between interior geostrophic modes and boundary Kelvin waves identified by the linear analysis performed in section 2:  $h=O\left(\mathrm{Ro}\right)$. {\color{black} Note that for a finite depth ocean, this would require $N \sim 1/\mathrm{Ro}$ layers.}

As in section 2, the flow domain takes place in a semi-infinite horizontal domain, with fields vanishing at infinity, and an impermeability constraint at the wall:
\begin{equation}
u_i(0,y,t)=0 \ .\label{eq:ucoast}
\end{equation}

We assume that the initial flow satisfies  quasigeostrophic scaling, with horizontal scale, vertical scale and velocities of order one, interface height variations of order $\mathrm{Ro}$, corresponding to vertical pressure variation between adjacent layers scaling as $\delta z$. The strategy is to divide the domain into an interior region satisfying standard quasigeostrophic equations, and a boundary layer with typical thickness scaling as  $\mathrm{Ro}$.
\subsection{Quasigeostrophic dynamics in the interior}
The interior dynamics is derived from Eqs.  (\ref{eq:ui_start}-\ref{eq:vi_start}-\ref{eq:hi_start}-\ref{eq:hi_pi}) following standard procedure based on asymptotic expansion in a low $\mathrm{Ro}$ limit \citep{vallis2017atmospheric}, with the ansatz
\begin{equation}
\left[\mathbf{u}_i,p_i\right]=\left[\mathbf{u}_{i,0}^g,p_{i,0}^g\right] +\mathrm{Ro}\left[\mathbf{u}_{i,1}^g,p_{i,1}^g\right] +O\left(\mathrm{Ro^2}\right) \label{eq:exp_g}
\end{equation}
We also assume that typical vertical variations of the pressure fields (up to order one) between adjacent layers is of order $\delta z$, consistently with the assumption of an initial condition satisfying quasigeostrophic scaling. According to Eq. (\ref{eq:hi_pi}), interior interface height variations  scale linearly with $ \mathrm{Ro}$. 
\begin{itemize}
\item  \textit{At order $0$}, one gets geostrophic balance, and hydrostatic balance still holds: 
\begin{equation}
\mathbf{u}_{i,0}^g= \left(-\partial_y \psi_i ,\, \partial_x \psi_i \right),\quad \psi_i\equiv p_{i,0}^g , \quad \eta_{i,0}^g=-\frac{1}{\mathrm{Bu}}\frac{p_{i+1,0}^g-2p_{i,0}^g+p_{i-1,0}^g}{\delta z^2}\ . \label{eq:eq_geo}
\end{equation}
\item \textit{At order $1$} we recover quasigeostrophic dynamics
\begin{equation}
\partial_t q^g_i +\mathbf{u}_{i,0}^g\cdot{\nabla} q_i^g=0 ,\quad q_{i}^g\equiv \nabla^2 \psi_i +{\mathrm{Bu}}^{-1} \left(\delta z \right)^{-2} \left({\psi_{i+1}-2\psi_i+\psi_i}\right)\label{eq:qg_def} \ .
\end{equation}
\end{itemize}
At this stage, one can not integrate the dynamics in Eq. (\ref{eq:qg_def}) for two reasons, both of them related {\color{black} with} potential vorticity inversion: (i) the boundary condition for $\psi$ at the wall remains  unknown; (ii) one can not rule out a source of vorticity within the boundary layer that would affect the streamfunction outside the boundary layer. To address those two issues, it is necessary to dwell into Kelvin boundary layer dynamics.

\subsection{Kelvin wave dynamics in the boundary layer}

According to the  analysis of linearized hydrostatic Boussinesq dynamics in section 2,  the Kelvin boundary layer dynamics is expected to be confined in a region of size $\mathrm{Ro} L$ away from the wall, with vertical variations of the fields taking place over a distance $\mathrm{Ro} H$. This motivates the following change of variable:
\begin{equation}
 X=\frac{x}{\mathrm{Ro}},\quad  \quad \delta Z = \frac{\delta  z}{\mathrm{Ro}}.
\end{equation}
The velocity and pressure fields in the boundary layer are decomposed as follows:
\begin{alignat}{7}
v_i &=&v_{i}^b(X,y,t)&+&v_{i,0}^g\big|_{x=0}(y,t)&+&\mathrm{Ro}v_{i,1}^g\big|_{x=0}(y,t)&+&\mathrm{Ro} X \partial_x v_{i,0}^g\big|_{x=0}(y,t)&+&O\left(\mathrm{Ro}^2\right),\label{eq:bound1}\\
u_i &=&u_{i}^b(X,y,t)&+&u_{i,0}^g\big|_{x=0}(y,t)&+&\mathrm{Ro}u_{i,1}^g\big|_{x=0}(y,t)&+&\mathrm{Ro} X \partial_x  u_{i,0}^g\big|_{x=0} (y,t)&+&O\left(\mathrm{Ro}^2\right),\label{eq:bound2}\\
p_i&=&p_{i}^b(X,y,t)&+& p_{i,0}^g |_{x=0}(y,t)&+& \mathrm{Ro}p_{i,1}^g |_{x=0}(y,t)&+&\mathrm{Ro} X \partial_x p_{i,0}^g|_{x=0}(y,t)&+&O\left(\mathrm{Ro}^2\right).\label{eq:bound3}
\end{alignat}
The matching condition between inner (index "b" for boundary) and outer (interior quasigeostrophic) solution is
\begin{equation}
\lim_{X\rightarrow +\infty} \left[u_i^b,v_i^b,p_i^b\right]=\left[0,0,0\right] \ . \label{eq:matching}
\end{equation}
The boundary fields are also expanded as
\begin{alignat}{6}
\left[\mathbf{u}_i^b,p_i^b\right]&=&\left[\mathbf{u}_{i,0}^b,p_{i,0}^b\right] &+&\mathrm{Ro}\left[\mathbf{u}_{i,1}^b,p_{i,1}^b\right]&+&O\left(\mathrm{Ro^2}\right)&,&\label{eq:exp_b}
\end{alignat}
and it will be convenient to decompose the total velocity and pressure fields as 
\begin{alignat}{6}
\left[\mathbf{u}_{i},p_{i}\right]&=& \left[\mathbf{u}_{i,0},p_{i,0}\right] &+&\mathrm{Ro}\left[\mathbf{u}_{i,1},p_{i,1}\right]&+&O\left(\mathrm{Ro^2}\right)\label{eq:exp_tot}&.&
\end{alignat}
The fields $\left[\mathbf{u}_{i,0},p_{i,0}\right]$ and $\left[\mathbf{u}_{i,1},p_{i,1}\right]$ include the trace of the interior field in the boundary layer regions as defined in Eq. (\ref{eq:bound1}-\ref{eq:bound2}-\ref{eq:bound3}). 

Special care must be taken to evaluate the different terms in the expansion of interface height variations  $\delta \eta_i$ defined in Eq. (\ref{eq:hi_pi}). Indeed, we have assumed that vertical variations of interior pressure  between adjacent layers scale as $\delta z$, and, based on linear analysis, we anticipated that vertical variations of boundary pressure  between adjacent layers scale as  $\delta Z$. Thus, the interface height variations can be expressed as  
\begin{equation}
\delta \eta_i=\delta \eta_{i,0} +O\left(\mathrm{Ro}\right),\quad  \delta \eta_{i,0}=-\frac{1}{\mathrm{Bu}}  \frac{p_{i+1,1}^b-2p_{i,1}^b+p_{i-1,1}^b}{\left(\delta Z\right)^2} \ .\label{eq:hi_piZ}
\end{equation}
Now that we have introduced the ansatz for the solution in the boundary layer, we write down the rescaled dynamical equations. Momentum equations   read 
\begin{alignat}{5}
\mathrm{Ro}\left( \partial_t +\mathrm{Ro}^{-1} u_i\partial_X +v_i\partial_y  \right)u_i   &=&-\mathrm{Ro}^{-1} \partial_X p_i &+&v_i&, \label{eq:u_dyn_bound}\\
  \mathrm{Ro}\left( \partial_t +\mathrm{Ro}^{-1}u_i\partial_X +v_i\partial_y  \right)v_i    &=&-\partial_y p_i &-& u_i & \ . \label{eq:v_dyn_bound}
  \end{alignat}
It will be convenient to use potential vorticity as a third dynamical equation:
\begin{equation}
    \left( \partial_t +\mathrm{Ro}^{-1}u_i\partial_X +v_i\partial_y  \right)q_i  =0,\quad q_i=\frac{1+\zeta_i}{1+\delta \eta_i} ,\quad 
    \zeta_i \equiv \partial_{X} v_i- \mathrm{Ro}\partial_y u_i\ .
\end{equation}
Consistently with the assumption of an initial condition satisfying quasigeostrophic scaling,
material conservation of potential vorticity for a fluid particle with initial relative vorticity $\zeta_i^{(t=0)}\sim \mathrm{Ro}$ and initial interface thickness variation $\delta \eta_i^{(t=0)}\sim \mathrm{Ro}$ can be recast as
\begin{equation}
    \zeta_i -\delta \eta_i=O\left(\mathrm{Ro} \right)  \ . \label{eq:pv_conservation_initial} 
\end{equation}

We inject the ansatz (\ref{eq:bound1}-\ref{eq:bound2}-\ref{eq:bound3}-\ref{eq:exp_g}-\ref{eq:exp_b}) in the rescaled dynamical system (\ref{eq:u_dyn_bound}-\ref{eq:v_dyn_bound}-\ref{eq:pv_conservation_initial}) and collect terms at each order with respect to $\mathrm{Ro}$. 
\begin{itemize}
    \item \textit{At order -1,} the momentum equation in $X$-direction yields %
    \begin{equation}
        \partial_X p_{i,0}=0\label{eq:order-1}
    \end{equation}
\item  \textit{At order 0,} the momentum and  potential vorticity equations yield respectively 
\begin{eqnarray}
 u_{i,0} \partial_X u_{i,0} &=&-\partial_X p_{i,1} +v_{i,0}, \label{eq:almost_geostrophic}\\
  u_{i,0} \partial_X v_{i,0}   &=&-\partial_y p_{i,0}-u_{i,0},  \label{eq:order0_2}\\
   \partial_X v_{i,0} -\delta \eta_{i,0} &=&0.\label{eq:pv_bound_order0}
 \end{eqnarray}
{\color{black} Differentiating} Eq. (\ref{eq:order0_2}) by $X$ and using Eq. (\ref{eq:order-1}) {\color{black} leads to  $ \partial_X\left(u_{i,0} \left(\partial_X v_{i,0}+1 \right)\right)=0$.  Using the  impermeability condition (\ref{eq:ucoast})  leads then to  $u_{i,0} \left(\partial_X v_{i,0}+1 \right)=0$} for all $X$. The case $\partial_X v_i=-1$  corresponds to a vanishing interface thickness, i.e. $1+\delta \eta_{i,0}=0$, according to Eq. (\ref{eq:pv_bound_order0}). This may occur along shock lines. 
From now on, we describe the flow dynamics away from these  isolated singularities. This corresponds to the second case $u_{i,0}(X,y,t)=0$. Using the matching condition (\ref{eq:matching}), we find an impermeability condition  for the geostrophic (interior) velocity field, and  a vanishing  across-wall velocity in the boundary layer:
\begin{equation}
    u_{i,0}^{g}\big|_{x=0}=0,\quad  u_{i,0}^{b}(X,y,t)=0 . \label{eq:v0b_geostrophic}
\end{equation} 
Eq. (\ref{eq:order0_2}) is now further simplified as $\partial_y p_{i,0}=0$. Using this equation  together with Eq. (\ref{eq:order-1}) and the matching condition (\ref{eq:matching}) yields 
\begin{equation}
    \quad p_{i,0}^b=0,\quad p_{i,0}^g\big|_{x=0}(y,t)=\psi_{i,\mathrm{wall}}(t).\label{eq:psi_wall}
\end{equation}
The second equality is the standard impermeability condition for quasigeostrophic flows along a wall. The value of $\psi_{i,\mathrm{wall}}$ will be determined using layerwise global mass conservation later on. Finally, Eq. (\ref{eq:almost_geostrophic})  and  (\ref{eq:pv_bound_order0}) boil down to 
\begin{equation}
    v_{i,0}^b=\partial_X p_{i,1}^b , \quad \frac{\partial^2}{\partial X^2}{p_{i,1}^b}= -\frac{1}{\mathrm{Bu}}  \frac{p_{i+1,1}^b-2p_{i,1}^b+p_{i-1,1}^b}{\left(\delta Z\right)^2} .\label{eq:Kelvin_rel}
\end{equation}
This shows that the triplet of boundary layer fields $[\mathbf{u}_{i,0}^b, p_{i,0}^b]$ satisfies the polarization relation of coastal Kelvin waves, as in Eq. (\ref{eq:kelvin_lin}). The boundary fields are then fully prescribed by the amplitude of $v_i^b$ at the boundary $X=0$. Their dynamics is obtained at next order. \item \textit{ At order 1,} the momentum equation in the $y$-direction evaluated at the wall yields 
\begin{eqnarray}
 \text{At $X=0$: }
 \partial_t v_{i,0} +  \partial_y\left( \frac{1}{2}v_{i,0}^2+p_{i,1}\right) &=&0  .\label{eq:bound_dyn_final}
\end{eqnarray}
We have used $u_{i}(0,y,t)=0$ and the order-$1$ impermeability constraint. Eq. (\ref{eq:bound_dyn_final}) can be recast  as a dynamical evolution for  $v_{i,0}^b(0,y,t)=v_{i,0}(0,y,t)-v_{i,0}^g(0,y,t)$, assuming that geostrophic fields are known. Noticing that $\partial_y p_{i,1}|_{X=0}= \partial_y p_{i,1}^b|_{X=0}(y,t) $, the combination of Eq. (\ref{eq:Kelvin_rel}) with   Eq.  (\ref{eq:bound_dyn_final}) and boundary condition (\ref{eq:matching}) provide  the  system of equations derived in \cite{deremble2017coupled}.
\end{itemize}

\subsection{Potential vorticity production by shallow water shocks}

\cite{dewar2010topographic,hogg2011kelvin,deremble2017coupled} showed that the boundary layer dynamics lead to shocks and the concomittant creation of cyclonic vorticity. Based on global conservation of circulation, \cite{deremble2017coupled} proposed a model for the feedback of these shocks on the interior quasigeostrophic dynamics. We propose here a more local justification of their model, relying on the theory of rotating shallow water shocks \citep{peregrine1998surf,pratt2007rotating,zeitlin2018geophysical}.

 A shallow water shock line {\color{black}, in the $X$-direction} indexed by $s$ in layer $i$ and located at $y=y_{s,i}(t)$ is associated with a jump of {\color{black} the} Bernoulli potential \textit{across} the shock{\color{black}, see e.g.   \citep{zeitlin2018geophysical}}:
\begin{equation}
    [ B_i]\equiv B_i(X,y_{s,i}^+)-B_i(X,y_{s,i}^-),\quad B_i(X,y_{s,i}^{\pm}) \equiv \frac{\mathbf{u}_i^2(X,y_{s,i}^{\pm})}{2}+p_i(X,y_{s,i}^{\pm}) \ .\label{eq:B_def}
\end{equation} 
When the value of $[B_i]$ varies \textit{along} the shock, {\color{black} in the $X$ direction,}there is a jump of potential vorticity \textit{across} the shock {\color{black} \citep{zeitlin2018geophysical}}:
\begin{equation}
    [q_i]\equiv q_i(X,y_{s,i}^+)-q_i(X,y_{s,i}^-) =\mathrm{Ro}^{-1}\frac{\partial_X \left([B_i] -\dot{y}_{s,i} [v_i]\right)}{h_i \left(v_i-\dot{y}_{s,i} \right) },
\end{equation}
where $[v_i]=v_i(X,y_{s,i}^+,t)-v_i(X,y_{s,i}^-,t)$ is the velocity jump across the shock, $\dot{y}_{s,i}\equiv \mathrm{d}{y}_{s,i}/\mathrm{d} t $ is the shock velocity  and $h_i(v_i-\dot{y}_{s,i})$ is the mass flux through the shock for an observer moving with the shock. This mass flux is conserved across the shock, with $[h_i(v_i-\dot{y}_{s,i})]=0$.
 The combination of a potential vorticity jump and a constant mass flux through the shock implies a net production of potential vorticity per unit time and per unit shock length, see e.g.  \citep{zeitlin2018geophysical}.  The total amount of potential vorticity production at $y=y_{s,i}$ in the boundary layer region is thus \begin{equation}
\int_0^{+\infty} \mathrm{d} X\ \partial_X \left([B_i] - \dot{y}_{s,i} [v_i]\right)=\dot{y}_{s,i}[v_i]_{X=0}-[B_i]_{X=0} ,
\label{eq:total_amount_pv}
\end{equation}
where the r.h.s. is the jump evaluated at $(X=0,y=y_{s,i})$. We have used the fact that there is no shock in the (quasigeostrophic) interior, for $X \rightarrow +\infty$. 

The net production of potential vorticity in the boundary layer contradicts our assumption of materially conserved potential vorticity used to derive the Kelvin wave dynamics in the boundary layer.
 One way to have a self-consistent model taking into account the local inviscid  production of vorticity  at $y=y_{s,i}$ is to inject in Eq. (\ref{eq:qg_def}) the total amount of potential vorticity of Eq. (\ref{eq:total_amount_pv}), at a distance $x=\sqrt{\mathrm{Ro}}$ much larger than the boundary layer of size $\mathrm{Ro}$, while remaining asymptotically  close to the wall: 
\begin{equation}
  \partial_t q_i^g+\mathbf{u}_i^g \cdot \nabla q_i^g= \delta(x-\sqrt{\mathrm{Ro}} )\delta(y-y_{s,i})  \left( \dot{y}_{s,i} [v_i]_{X=0}-[B_i]_{X=0}\right). \label{eq:pv_qg_modified}
  \end{equation}
 {\color{black} This  infinitesimal shift of  potential vorticity production from the boundary layer to the interior region  is the only phenomenological step of the model derivation. It is motivated by numerical simulations showing production of cyclones through the detachment of boundary layers close to the shock location in primitive equation models \citep{deremble2017coupled}. This can also be interpreted as the continuous version of the discrete numerical algorithm used by \citep{deremble2017coupled} to simulate their coupled reduced model: for a given grid size, the location of the source term at $x=\sqrt{\mathrm{Ro}}$ guarantees that potential vorticity injection occurs within the cell adjacent to the wall in the limit $\mathrm{Ro} \rightarrow 0$.}
 
 To be consistent with this procedure of potential vorticity injection in the interior following the formation of shocks in the boundary, the total circulation in the boundary regions must be left invariant, which, assuming that it is initially zero, implies  
\begin{equation}
\Gamma_i=\Gamma_i^g,\quad  \text{with }   \Gamma_i\equiv -\int_{-\infty}^{+\infty} \mathrm{d} y \ v_{i,0},\quad \Gamma_i^g\equiv -\int_{-\infty}^{+\infty} \mathrm{d} y \ v_{i,0}^g .\label{eq:def_circulation}
\end{equation}

\subsection{Mass conservation, quasigeostrophic circulation, and final set of equations}

The full dynamical system coupling boundary dynamics with quasigeostrophic interior is yet not closed, as one still must determine the value of $\psi_{
i,\mathrm{wall}}$ introduced in Eq. (\ref{eq:psi_wall}). This is settled by using layerwise, global mass conservation: 
\begin{equation}
    \left<\delta \eta_i \right>=0,\quad \text{with } \left<\delta \eta_i \right>\equiv\int_{0}^{+\infty} \mathrm{d} x \int_{-\infty}^{+\infty} \mathrm{d} y \ \delta \eta_i .\label{eq:def_eta}
    \end{equation}
    The  difficulty with respect to classical quasigeostrophic models is that variations of mass in the boundary layers are of the same order as variation of mass in the interior. Despite this subtlety, the use of Eq. (\ref{eq:def_circulation}) allows us to recover the constraint  (see Appendix B):
\begin{equation}
\left<\psi_{i+1}+\psi_{i-1}-2\psi_i\right>=0 .
 \label{eq:mass_cond}
\end{equation}
The set of boundary values $\psi_{i,\mathrm{wall}}$ is deduced from the set of constraints  in (\ref{eq:mass_cond}), following standard procedure \citep{mcwilliams1977note}.  
Let us note that Eq. (\ref{eq:mass_cond}) implies instantaneous adjustment of the mass in each interior layer. The reasons is that we assumed previously that quasigeostrophic motion has typical vertical scale of order $1$ (size $H$ in dimensional units), and that Kelvin waves associated with vertical variations of order $1$ are filtered out in the asymptotic expansion.  {\color{black}Note that our phenomenological procedure of potential vorticity injection is such that structures of vertical size much smaller than $H$ can be formed in the interior, depending on shock properties in the boundary layer. This injection procedure induces therefore an inconsistency with respect to the initial hypothesis on the size of interior flows. By applying mass conservation (\ref{eq:mass_cond}), we continue to assume instantaneous adjustment for those smaller scale structures in the interior.  Possible resonances between slow Kelvin wave dynamics and interior quasi-geostrophic flow is still taken into account through the boundary layer equation (\ref{eq:bound_dyn_final}).}

Finally, the full coupled system is given by potential vorticity advection in Eq.  (\ref{eq:pv_qg_modified}) and Kelvin wave dynamics in Eq. (\ref{eq:bound_dyn_final}). The interior velocity field is obtained by inversion of the quasigeostrophic potential vorticity field defined in Eq. (\ref{eq:qg_def}), using the lateral boundary conditions in Eq. (\ref{eq:psi_wall}) and the constraints (\ref{eq:mass_cond}). Kelvin wave dynamics in Eq. (\ref{eq:bound_dyn_final}) depends on the geostrophic interior field evaluated at the boundary; in turn, Eq. (\ref{eq:bound_dyn_final}) is used to find shock locations and evaluate the corresponding  velocity and Bernoulli potential jumps appearing in the r.h.s. of Eq.   (\ref{eq:pv_qg_modified}), and defined in Eq. (\ref{eq:B_def}). {\color{black} Thus, the knowledge of potential vorticity production is bound to the knowledge of Bernoulli potential jumps across shocks in the boundary layer dynamics. We have up to now not explained how to determine the actual value of such Bernoulli potential jumps. In the case of a one-layer shallow water model, one just needs to apply standard local mass and momentum conservation across the shock. However, the problem is indeterminate in the case of multiple-layer  shallow water flows, and no universal rule exist \citep{zeitlin2018geophysical}. One then either needs  to introduce additional phenomenological assumptions on the shock behavior, or to bypass this issue by regularising the boundary Kelvin wave dynamics with dissipative terms. In the latter case, the  Bernoulli potential jumps is estimated numerically across quasi-shocks that are defined at locations where gradients exceed a given threshold; this is the approach followed in \citep{deremble2017coupled}, who introduced viscous dissipation in Eq. (\ref{eq:bound_dyn_final}).}

\subsection{Conservation of quasigeostrophic energy}

{\color{black} Let us now consider the case of a finite-depth ocean with $N$ layers of thickness $h$, such that $Nh=1$, with a rigid lid approximation (at layer $i=N$) and flat bottom boundary condition (at layer $i=1$).  The dynamics is  the same as in the case of an  infinite number of layers, with two additional constraints related to upper and lower boundary conditions, namely $\psi_{N+1}=\psi_N$ and $\psi_0=\psi_1$, respectively.} The quasigeostrophic {\color{black} energy} is  defined as 
\begin{equation}
     E^g \equiv \frac{{\color{black} h}}{2} \sum_{i=1}^N\left< \left(\nabla \psi_i\right)^2 +\frac{1}{\mathrm{Bu}}\left(\frac{\psi_i-\psi_{i-1}}{\delta z}\right)^2\right>.
\end{equation}
We assume the presence of $n_{i}$ shocks in each layer $i$. The shocks are  indexed by $(s,i)$ with $ 1 \le s\le n_{i}$. Their location in the $y$ direction is denoted $y_{s,i}(t)$,  {\color{black}and the corresponding potential vorticity injection rate is $\gamma_{s,i}\equiv \dot{y}_{s,i}[v_i]-[B_i]$, see Eq. (\ref{eq:pv_qg_modified})}. The temporal evolution of quasigeostrophic energy is computed by using the dynamical equation (\ref{eq:pv_qg_modified}), the definition of quasigeostrophic potential vorticity in Eq. (\ref{eq:qg_def}), as well as the definition of circulation in Eq. (\ref{eq:def_circulation}) and mass conservation in Eq. (\ref{eq:mass_cond}):
\begin{equation}
   \frac{\mathrm{d}}{\mathrm{d}t} E^g={\color{black} h }\sum_{i=1}^N\left( -\psi_{i,\mathrm{wall}}\frac{\mathrm{d}\Gamma_i^g}{\mathrm{d} t}+\sum_{s=1}^{n_{i}}  \gamma_{s,i}\psi_i(\sqrt{\mathrm{Ro}},y_{s,i},t)\right),\quad  \frac{\mathrm{d}}{\mathrm{d} t} \Gamma_i^g =\sum_{s=1}^{n_{i}}\gamma_{s,i}.
\end{equation}
{\color{black} Following the conventions used for the asymptotic analysis, the initial energy and the initial circulations are of order one; according to this asymptotic analysis, the potential vorticity injection rate $\gamma_{s,i}$ is also of order one. Using $hN\sim 1$ and $\psi_i(\sqrt{\mathrm{Ro}},y_{s,i})=\psi_{i,\mathrm{wall}}+\sqrt{Ro}\partial_x \psi|_{0,y_{s,i}} +o(\sqrt{\mathrm{Ro}})$, we find that the   total quasigeostrophic energy vanishes in the limit $\mathrm{Ro}\rightarrow 0$.
We conclude that quasigeostrophic energy is conserved, {\color{black}  unless the asymptotic approach fails in such a way that $\gamma_{s,i}$ scales as $1/\sqrt{\mathrm{Ro}}$.} This result does not contradict the observation of enhanced dissipation in the presence of a coast  \citep{deremble2017coupled}.  It just suggests that the amplitude of enhanced dissipation should tend to zero with $\mathrm{Ro}$, so that the corresponding energy sink in the actual ocean would be a finite-$\mathrm{Ro}$ effect. It will be interesting to investigate how enhanced dissipation actually scales with the $\mathrm{Ro}$ in numerical models.}

\section{Discussion and conclusion}

We have revisited the derivation of Deremble-Johnson-Dewar model coupling interior continuously stratified quasigeostrophic fluid to a boundary layer with low-frequency Kelvin wave dynamics. The boundary layer thickness  scales linearly with the Rossby number, and the dynamics inside this layer is described by a  two-dimensional dynamical equation  at the wall. This wall dynamics leads to shocks. Our contribution is to clarify the matching condition between interior and boundary dynamics through mass conservation and shock properties, and to show that quasigeostrophic energy is conserved:  shocks are an inviscid sink of energy, but those sinks are confined in a narrow boundary layer {\color{black} whose width scales linearly
with $\mathrm{Ro}$}, so that their net contribution vanishes in the small $\mathrm{Ro}$ limit. 

The original set of hydrostatic Boussinesq equations on the $f$-plane breaks time reversal symmetry. The symmetry breaking parameter is the Rossby number $\mathrm{Ro}$. The quasigeostrophic model on the unbounded $f$-plane is derived in the limit $\mathrm{Ro}\rightarrow 0$. the Rossby number is not a parameter of this reduced model. Time-reversal symmetry is thus recovered in $f$-plane quasigeostrophic equations.  The addition of a wall allows for the propagation of unidirectional Kelvin waves that bring back broken time-reversal symmetry into  quasigeostrophic dynamics. This broken symmetry manifests itself in the interior flow as the formation of quasigeostrophic cyclones along the coast by Kelvin wave shocks: just as surface boundary layers favour cyclonic structures \citep{roullet2010cyclone}, lateral Kelvin boundary layers break cyclone-anticyclone symmetry.

Cyclones injected at the boundary start to impact the anticyclonic interior flow when the vertically integrated interior anticyclonic circulations become of the same order as the total amount of injected potential vorticity. Since injection takes place over a vertical scale of order $\mathrm{Ro}$ with a circulation production rate of order one, the interaction time can be estimated as $T_{\mathrm{int}}\sim \mathrm{Ro}^{-1}$. The validity of the model in this long-time limit remains to be proven: shocks inject in the interior cyclonic structures with vertical size of order $\mathrm{Ro}$, which seems to contradict the initial assumption of  quasigeostrophic structures with order one vertical variations (scale $H$ in dimensional units). At a phenomenological level, one could argue that inverse cascade and barotropization processes organize the initially shallow cyclones into deeper ones. 

This paper focused on inviscid dynamics and thus left aside the role of viscous boundary layers. In two-dimensional turbulence, the detachment of these layers  may lead to dissipative structures  \citep{waidmann2018energy}, and drastically changes the interior vorticity dynamics \citep{roullet2020}. The role of viscous boundary layers  in continuously stratified rotating flows remains to be addressed. 

While quasigeostrophic energy remains a conserved quantity  at lowest order in $\mathrm{Ro}$, boundary layer Kelvin dynamics plays an active role on the interior flow patterns, through the injection of cyclonic vorticity close to the coast. This could be a key aspect of   oceanic western boundary currents detachment \citep{deremble2017coupled}. The $f$-plane coastal problem can also be interpreted as a toy model for the dynamics of equatorial planetary flows with symmetric temperature fields. Deremble-Johnson-Dewar mechanism could offer in this framework an explanation for the generation of intense equatorial cyclonic dipolar structures. Such patterns are an essential feature of Madden-Julian oscillations \citep{rostami2019eastward}.  For this reason, we think that Deremble-Johnson-Dewar  mechanism for the production of sub-mesoscale oceanic structures also deserves  attention in the context of equatorial atmospheric flows. 
{\color{black}
\section*{Acknowledgement}
This work has been motivated by a presentation in the working group GDT MathsInFluids organized by C. Lacave, D. Bresh and L. Saint-Raymond. A.V. warmly thank Bruno Deremble for fruitful discussions, Pierre Delplace for help with Mathematica, and the three reviewers who helped to improve significantly the presentation. Funding:  ANR-18-CE30-0002-01 WTF, and MSOM project. 
\section*{Declaration of interests}
The author reports no conflict of interest.
\section*{Appendix A: hydrostatic relations for multiple-layer models}

We consider an ocean model with $N$ fluid layers. 
The layers are indexed in the upward direction by $i$, with $1 \le i \le N$. We  assume a constant atmospheric pressure $P_a$ above the upper layer $i=N$. 
The total pressure fields, the thickness fields and the density fields in each layer are denoted $P_i(x,y,z,t)$, $h_i(x,y,t)$, $\rho_i=\rho_0+(N-i)\Delta \rho$, respectively. The interface elevation relative to a rest state and the interface depth between layers $i$ and $i+1$ are denoted $\eta_{i+0.5}$ and $z_{i+0.5}$. 
In dimensional units,  interface height elevations, interface depth, and interface thickness are related through 
\begin{equation}
z_{i+0.5}=(i-N) h +\eta_{i+0.5} \quad h_i = h\left(1+\delta \eta_i\right),\quad \delta \eta_i \equiv \frac{\eta_{i+0.5}-\eta_{i-0.5}}{h}.
\end{equation}
The pressure is deduced from hydrostatic balance:
\begin{equation}
    P_i -P_a= g\rho_i (z_{i+0.5}-z)+\sum_{j=i+1}^N g \rho_j h_j,
\end{equation}
 For $1<i<N$, a straightforward computation yields  to
\begin{equation}
    P_{i+1}-2P_i-P_{i-1} = -g h \Delta \rho \left(\delta \eta_i +1\right). \label{eq:p}
\end{equation}
We now introduce the rescaled dynamical pressure $p_i\equiv P_i/(\rho_0 UfL)+F_i(z)$. This field  is defined in each layer up to a function of $z$, since only its horizontal gradient matters in the dynamics. We choose the gauge function $F_i(z)$ in such a way to cancel the constant term in the r.h.s. of Eq. (\ref{eq:p}). Recalling that $g\Delta \rho h/(\rho_0 fUL)=(h/H)^2\mathrm{Bu}/\mathrm{Ro}$, we obtain Eq.  (\ref{eq:hi_pi}). This relation hold whatever the scaling of $N$ and $h$ with $\mathrm{Ro}$.

\section*{Appendix B: Global mass conservation}
Global conservation of mass in each layer $i$ is expressed in Eq. (\ref{eq:def_eta}). 
We decompose the integral in the $x$ direction as a boundary term and an interior term:
\begin{equation}
    \int_{0}^{+\infty}\mathrm{d} x\  \delta \eta  =\int_{0}^{\mathrm{Ro}^{3/4}} \mathrm{d} x \  \delta \eta +\int_{\mathrm{Ro}^{3/4}}^{+\infty}\mathrm{d} x \   \delta \eta  .
\end{equation} Changing variable in the first integral with  $X=x/\mathrm{Ro}$ and considering the small $\mathrm{Ro}$ limit yields, at  order $\mathrm{Ro}$:
\begin{equation}
 \int_{0}^{+\infty} \mathrm{d} X \int_{-\infty}^{+\infty} \mathrm{d} y \ \delta \eta_{i,0} + \int_{0}^{+\infty} \mathrm{d} x \int_{-\infty}^{+\infty} \mathrm{d} y  \ \delta \eta_{i,1}^g= 0,\label{eq:mass_inter1}
\end{equation}
where $\delta \eta_{i,0}$ is the order zero interface height variation in the boundary region and $\delta \eta_{i,1}^g$ is the order one interior geostrophic interface height variation  defined in Eq. (\ref{eq:eq_geo}). 
Variations of mass in the boundary layers are of the same order as in the interior since interface height variation $\delta \eta$ scales as $\mathrm{Ro}$ over a a region of size $1$ in the quasigeostrophic region, while $\delta \eta$ scales as $1$ over a region of size $\mathrm{Ro}$ in the boundary layer. Using the expression of $\eta_{i,0}^g$ in Eq. (\ref{eq:eq_geo}), injecting  Eq. (\ref{eq:pv_bound_order0})  in Eq. (\ref{eq:mass_inter1}), mass conservation reads
\begin{equation}
\int_0^{+\infty}\mathrm{d} x \int_{-\infty}^{+\infty}\mathrm{d} y \ \left(\psi_{i+1}+\psi_{i-1}-2\psi_i\right) =  \mathrm{Bu} {\delta z^2} \int_0^{+\infty}\mathrm{d} X \int_{-\infty}^{+\infty}\mathrm{d} y\  \partial_X v_{i,0} \ .\label{eq:mass_der}
\end{equation}
 Using $v_{i,0}(+\infty,y,t)=v^g_{i,0}(0,y,t)$, the definition of circulations $\Gamma_i,\Gamma_i^g$ in Eq. (\ref{eq:def_circulation}), and the notation  $\left< \psi_i\right>=\int_0^{+\infty}\mathrm{d} x \int_{-\infty}^{+\infty}\mathrm{d} y \ \psi_i$, integration of the r.h.s. in Eq. (\ref{eq:mass_der}) yields
\begin{equation}
 \left<\psi_{i+1}+\psi_{i-1}-2\psi_i\right> = \mathrm{Bu} \delta z^2\left( \Gamma_i-\Gamma_i^g\right) ,\label{eq:mass_der}
\end{equation}
The mass constraint in Eq. (\ref{eq:mass_cond}) follows from the equality $\Gamma_i=\Gamma_i^g$.} 
\bibliographystyle{jfm}

\bibliography{anomalous_qg}

\end{document}